\documentclass[prd,nofootinbib,floats,superscriptaddress,eqsecnum,tightenlines,11pt]{revtex4}

\usepackage{hyperref}
\usepackage{graphicx}
\usepackage{setspace}
\usepackage{color}
\usepackage{amsmath,amssymb,amsfonts,amsthm,latexsym,stmaryrd}

\def\be{\begin{equation}}
\def\ee{\end{equation}}
\def\ba{\begin{eqnarray}}
\def\ea{\end{eqnarray}}

\def\eps{\varepsilon}

\def\de{\mathrm{d}}

\def\f{\frac}
\def\lb{\big\lbrace}
\def\rb{\big\rbrace}
\def\SU{\text{SU}}
\def\SO{\text{SO}}

\def\su{\mathfrak{su}}
\def\so{\mathfrak{so}}

\def\time{\,{\scriptstyle{\times}}\,}

\begin{document}

\title{A note on the Holst action, the time gauge,\\and the Barbero-Immirzi parameter}

\author{Marc Geiller}
\affiliation{Institute for Gravitation and the Cosmos \& Physics Department, Penn State, University Park, PA 16802, U.S.A.}
\author{Karim Noui}
\affiliation{Laboratoire de Math\'ematiques et Physique Th\'eorique, Universit\'e Fran\c cois Rabelais, Parc de Grandmont, 37200 Tours, France}
\affiliation{Laboratoire APC -- Astroparticule et Cosmologie, Universit\'e Paris Diderot Paris 7, 75013 Paris, France}

\begin{abstract}
In this note, we review the canonical analysis of the Holst action in the time gauge, with a special emphasis on the Hamiltonian equations of motion and the fixation of the Lagrange multipliers. This enables us to identify at the Hamiltonian level the various components of the covariant torsion tensor, which have to be vanishing in order for the classical theory not to depend upon the Barbero-Immirzi parameter. We also introduce a formulation of three-dimensional gravity with an explicit phase space dependency on the Barbero-Immirzi parameter as a potential way to investigate its fate and relevance in the quantum theory.
\end{abstract}

\maketitle

\section*{Introduction}

\noindent The Barbero-Immirzi parameter \cite{barbero,immirzi} $\gamma$ plays a rather intriguing role in loop quantum gravity. In the classical theory, this role can be analyzed both at the Lagrangian and at the Hamiltonian level. In the former case, the Barbero-Immirzi parameter is introduced as the coupling constant of a topological\footnote{Strictly speaking, the Holst term is not a topological term in the usual sense because it cannot be written as the exterior derivative of a three-form, unlike the truly topological Pontrjagin, Euler, and Nieh-Yan terms. What we mean here is simply that the Holst term does not affect the classical theory.} term that is added to the first order Hilbert-Palatini action in order to obtain the Holst action \cite{holst}. This topological term being vanishing once we resolve the torsion-free equations of motion in order to expresses the spin connection in terms of the tetrad, the Barbero-Immirzi parameter drops out of the classical theory in the absence of torsion. It is clear that this mechanism is intimately related to the dynamics of the theory, since some of the equations that are necessary in order to express the spin connection in terms of the tetrad are dynamical equations. In contrast, the Hamiltonian formalism contains some subtleties, which are related to the use of the time gauge and the introduction of the $\su(2)$ Ashtekar-Barbero connection. Indeed, once the second class constraints of the Holst action in the time gauge are solved, the resulting Hamiltonian contains an explicit dependency on $\gamma$, and it is non-trivial to identify the reason for which this dependency disappears at the dynamical level. This is to be contrasted with the canonical analysis of the Holst action without the time gauge \cite{alexandrov1,alexandrov2,alexandrov3,alexandrov4}, in which the reduced phase space given by the Dirac brackets contains no dependency on $\gamma$. Interestingly, this observation is also true at the discrete level, and it has been shown in \cite{dittrich-ryan1,dittrich-ryan2} that depending on wether all or only a subset of the discrete simplicity constraints are imposed, the resulting phase space is either the $\gamma$-dependent one of twisted geometries, or the $\gamma$-independent one of Regge geometries.

In the quantum theory, the Barbero-Immirzi parameter appears in the spectrum of the kinematical geometrical operators \cite{smolin-area,rovelli-smolin-area-volume,ashtekar-lewandowski-area,ashtekar-lewandowski-volume}, it shows up in the black hole entropy formula \cite{rovelli-black hole,ABK,meissner,agullo,ENP} (although it was argued recently that this dependency could be removed \cite{gosh-perez}, and it is now possible to compute the entropy for $\gamma=\pm i$ \cite{usBH}), and it plays a crucial role in the definition of the spin foam amplitudes for the quantum dynamics \cite{EPR,Engle:2007wy,livine-speziale2,Freidel:2007py}. Many people have discussed its significance and suggested alternative physical interpretations \cite{rovelli-thiemann,mercuri-gamma,mercuri-gamma2,mercuri-randono,mercuri-taveras,taveras-yunes,menamarugan}, and its status is still quite unclear. In \cite{BS}, it was shown in the context of perturbative first order gravity that when the Euclidean theory is coupled to fermions the value $\gamma^2=1$ is a UV fixed point. This comes as another indication that the (here Euclidean) self-dual value plays a particular role in quantum gravity. It has furthermore been argued \cite{FS} that $\gamma$ measure the strength of the torsion fluctuations in the path integral of quantum gravity, and that any value different form $\gamma\in\{0,\infty\}$ leads to CP violations.

It is well known that if one performs the canonical analysis of the first order Hilbert-Palatini action, the resulting phase space is that of the second order ADM formulation of gravity \cite{peldan,romano}. Using the Holst action and the time gauge to construct the Ashtekar-Barbero connection is the only way to obtain, at the end of the canonical analysis, a phase space parametrized by a gauge connection and its conjugate momentum, i.e. a first order formulation\footnote{The complex formulation in terms of the (anti) self-dual Ashtekar connection is also first order, but it has to be supplemented by the reality conditions ${}^{\mathbb{C}\!\!}\bar{A}^i_a+{}^{\mathbb{C}\!\!}A^i_a=\Gamma^i_a(E)$.}. The question is therefore wether at the dynamical level the torsion-free equations of motion are recovered, and if yes wether this indeed implies the disappearance of the Barbero-Immirzi parameter. In this note, we identify at the Hamiltonian level (for the Holst action in the time gauge) the various components of the covariant torsion two-form $T^I_{\mu\nu}$, and show that they are all vanishing for various reasons: $T^i_{ab}$ is vanishing because of the resolution of the second class constraints; $T^0_{ab}$ is vanishing on the surface of the Gauss and second class constraints; $T^0_{0a}$ is vanishing as a consistency condition on the Lagrange multipliers; and finally $T^i_{0a}$ is vanishing once the dynamics of the triad is computed.

At the level of the classical dynamics, one can therefore expect that if all these components are forced to vanish (by computing the dynamics of the electric field and fixing the Lagrange multipliers), the physical observables will be independent of $\gamma$. However, because of the quantum fluctuations, one can still expect a $\gamma$-dependency in the quantum theory, even after taking into account the dynamics. This is just a consequence of the fact that some torsion-free equations will not be imposed strongly in the quantum theory. In fact, beyond these considerations on the vanishing of the torsion, and even regardless of all the quantization ambiguities that are involved in the definition of the quantum dynamics of loop quantum gravity, it is known that the choice of variables at the classical level might be at the origin of an anomaly. The reason for this is that the Ashtekar-Barbero connection is not a spacetime connection, i.e. it does not transform as a one-form under the action of the total Hamiltonian generating time evolution. This was originally pointed out by Samuel on a simple example \cite{samuel}, and later on proven by Alexandrov at the level of the Lorentz-covariant formulation \cite{alexandrov5}. Here we re-derive this result in the case of the time gauge Holst action. This lack of spacetime interpretation of the Ashtekar-Barbero connection potentially has far reaching consequences, since even if the Hamiltonian torsion equations are imposed at the dynamical level, the anomalous transformation of the connection under time evolution will introduce an extra dependency on $\gamma$.

Since we are only able to consider the classical theory and not investigate precisely the fate of the Barbero-Immirzi parameter at the quantum level, it would be nice to have an exactly solvable model that bears enough similarities with the four-dimensional Ashtekar-Barbero theory and that can be used as a testbed. We achieve this note by proposing such a model, which is nothing else than three-dimensional gravity with a Barbero-Immirzi parameter. Based on the action for three-dimensional gravity introduced in \cite{GN}, we show that two different gauge choices lead to two different parametrization of the phase space: one that is $\gamma$-independent, and one that has the exact same structure as the four-dimensional Ashtekar-Barbero theory, along with its $\gamma$-dependency. This is very advantageous since the quantization of three-dimensional gravity is a well-known topic that has been studied from many points of view.

This note is organized as follows. In the first section, we study the four-dimensional Holst theory. After reviewing its Lagrangian formulation, we perform the canonical analysis in the time gauge in a manner more transparent than the original study of \cite{holst}. We focus in particular on the components of the torsion two-form and show by which mechanisms they vanish at the Hamiltonian level. Then we briefly recall the algebra of constraints of the theory and give the generators of the gauge symmetries. This enables us to compute the time evolution given by the total Hamiltonian, and in particular to show that, apart from the case of the self-dual theory $\gamma=\pm1$, the Ashtekar-Barbero connection does not transform as a one-form under time diffeomorphisms. In the second section, we introduce the formulation of three-dimensional gravity with a Barbero-Immirzi parameter.

Our notations are such that $\mu,\nu,\dots$ refer to spacetime indices, $a,b,\dots$ to spatial indices, $I,J,\dots$ to $\so(4)$ indices, and $i,j,\dots$ to $\su(2)$ indices. We assume that the $d$-dimensional spacetime manifold $\mathcal{M}$ is topologically $\Sigma\times\mathbb{R}$, where $\Sigma$ is a $(d-1)$-dimensional manifold without boundaries. We use the notation $u^I$ for vectors in $\mathbb{R}^4$ with components $(u^0,u^i)$, the cross product between elements $u,v\in\mathbb{R}^3$ to denote the operation $(u\time v)^i=\eps^{i}_{~jk}u^jv^k$, and the dot product for $u\cdot v=u^iv_i$. We often denote the vectors $u^i\in\mathbb{R}^3$ simply by $u$. Symmetrization and anti-symmetrization of indices are defined with weights $1/2$. We will focus exclusively on the Euclidean case, but of course the results extend to the Lorentzian theory as well.

\section{The Holst action for gravity}
\label{sec1}

\noindent Let us start by recalling some well known facts about the first order Hilbert-Palatini action for gravity. In terms of the spacetime $\so(4)$ gauge connection $\omega^{IJ}_\mu$ and the tetrad $e^I_\mu$, it is given by
\be\label{HP lagrangian}
S_\text{HP}=\f{1}{8}\int_\mathcal{M}\eps_{IJKL}e^I\wedge e^J\wedge F^{KL}=\frac{1}{8}\int_\mathcal{M}\de^4x\,\eps^{\mu\nu\rho\sigma}\eps_{IJKL}e_\mu^Ie_\nu^J 
F_{\rho\sigma}^{KL},
\ee
where $F_{\mu \nu}^{IJ}=\partial_\mu\omega_\nu^{IJ}-\partial_\nu\omega_\mu^{IJ}+[\omega_\mu,\omega_\nu]^{IJ}$ denotes the curvature of the connection. The equations of motion can be derived as follows. Using the so-called Palatini equation $\delta F^{IJ}_{\mu\nu}=2D_{[\mu}^{\phantom{I}}\delta\omega^{IJ}_{\nu]}$ and an integration by parts, we first obtain
\be
\f{\delta S_\text{HP}}{\delta\omega^{IJ}_\mu}=0\quad\Rightarrow\quad\eps^{\mu\nu\rho\sigma}\eps_{IJKL}D_\nu\big(e^K_\rho e^L_\sigma\big)=0.
\ee
Assuming that the co-tetrad is invertible, the meaning of this equation is that the torsion
\be
T^I_{\mu\nu}\equiv\partial_\mu e^I_\nu-\partial_\nu e^I_\mu+\omega^I_{\mu J}e^J_\nu-\omega^I_{\nu J}e^J_\mu
\ee
is vanishing (in the presence of matter sources, this is not true anymore). The torsion-free condition can then be solved for the connection, leading to the Levi-Civita spin connection
\be\label{torsion-free spin}
\omega^{IJ}_\mu(e)=-e^{\nu J}\partial_\mu e^I_\nu+e^{\nu J}e^I_\sigma\Gamma^\sigma_{~\mu\nu}(g)
=\f{1}{2}e^{\sigma[I}\left(\partial_{[\mu}^{\phantom{J]}}\!e_{\sigma]}^{J]}+e^{\rho J]}e^L_\mu\partial_\rho e_{\sigma L}\right).
\ee
Varying the action with respect to the tetrad, we obtain the second set of equations of motion, namely
\be
\f{\delta S_\text{HP}}{\delta e^I_\mu}=0\quad\Rightarrow\quad\eps^{\mu\nu\rho\sigma}\eps_{IJKL}e^J_\nu F^{KL}_{\rho\sigma}=0,
\ee
which, upon use of the solution \eqref{torsion-free spin} to specify $F^{IJ}_{\mu\nu}[\omega]=F^{IJ}_{\mu\nu}[\omega(e)]$, become simply the vacuum Einstein equations $G_{\mu\nu}=0$.

The Hilbert-Palatini Lagrangian is not the only locally Lorentz-invariant four-form that can be constructed with the tetrad and the spin connection. One can indeed add to it a cosmological term, as well as several topological terms \cite{rezende-perez}. These are the Pontryagin invariant, the Euler invariant, the Nieh-Yan invariant, and the Holst invariant. The action corresponding to the Holst term is
\be
S_\text{H}=\f{1}{4}\int_\mathcal{M}\delta_{IJKL}e^I\wedge e^J\wedge F^{KL},
\ee
where $\delta_{IJKL}\equiv\delta_{I[K}\delta_{L]J}$. Using the Levi-Civita spin connection $\omega^{IJ}_\mu(e)$ defined by \eqref{torsion-free spin}, this term is clearly irrelevant since it just amounts to writing the Bianchi identity
\be
\eps^{\mu\nu\rho\sigma}\delta_{IJKL}e^I_\mu e^J_\nu F^{KL}_{\rho\sigma}[\omega(e)]\equiv\eps^{\mu\nu\rho\sigma}R_{\mu\nu\rho\sigma}[\Gamma(g)]=0.
\ee
The Holst term is therefore vanishing on-shell, i.e. when the torsion free condition is enforced.

The action that we are interested in is given by a combination of the Hilbert-Palatini action with the Holst invariant, where this latter is multiplied by a coupling constant $\gamma\in\mathbb{R}-\{0\}$ known as the Barbero-Immirzi parameter. This gives the Holst action
\be\label{holst action}
S\equiv S_\text{HP}+\gamma^{-1}S_\text{H}=\frac{1}{4}\int_\mathcal{M}\de^4x\,\eps^{\mu\nu\rho\sigma}\left(\frac{1}{2}\eps_{IJKL}e_\mu^Ie_\nu^J 
F_{\rho\sigma}^{KL}+\gamma^{-1}e_\mu^Ie_\nu^JF_{\rho \sigma}^{IJ}\right).
\ee
Although it plays no role at all at the classical level since it multiplies a term which is vanishing on-shell, the Barbero-Immirzi parameter does not disappear at the quantum level. For this reason, it cannot be set to an arbitrary value in the Holst action if we are interested in studying the quantum theory. Also, when coupling gravity to fermions, the dependence on $\gamma$ becomes nontrivial, and wether this might lead to potentially observable effects has been debated quite a lot \cite{fermions1,fermions2,fermions3,fermions4,fermions5}.

\subsection{Hamiltonian analysis in the time gauge}

\noindent To perform the canonical analysis of the Holst action \eqref{holst action}, it is necessary to split the spacetime and internal indices into their various components. For the tetrad field, this decomposition is given by
\be\label{tetrad decomposition}
\begin{tabular}{lllllll}
$e^0_\mu\de x^\mu$ & = & $e^0_0\de x^0+e^0_a\de x^a$ &~~~~~~~~~~~~& $e^i_\mu\de x^\mu$ & = & $e^i_0\de x^0+e^i_a\de x^a$ \cr
	  & $\equiv$  & $N\de x^0+\chi_ie^i_a\de x^a$,        &~~~~~~~~~~~~&            & $\equiv$ & $N^i\de x^0+e^i_a\de x^a$,
\end{tabular}
\ee
and we can further write $N^i\equiv N^ae^i_a$. From now on, we are going to work with the time gauge, which corresponds to setting $\chi_i=0$. Let us first write the various components of the covariant torsion $T^I_{\mu\nu}$. In the time gauge, they are given by
\begin{subequations}\label{torsion components}
\ba
T^i_{ab}&=&\partial_ae^i_b-\partial_be^i_a+\omega^{ij}_ae^j_b-\omega^{ij}_be^j_a+\omega^{i0}_ae^0_b-\omega^{i0}_be^0_a\\
&=&\partial_ae^i_b-\partial_be^i_a-\eps^i_{~jk}(\omega_a^je^k_b-\omega_b^je^k_a),\label{torsioniab}\\
T^0_{ab}&=&\partial_ae^0_b-\partial_be^0_a+\omega^{0i}_ae^i_b-\omega^{0i}_be^i_a\\
&=&\omega^{0i}_ae^i_b-\omega^{0i}_be^i_a,\label{torsion0ab}\\
T^0_{0a}&=&\partial_0e^0_a-\partial_ae^0_0+\omega^{0i}_0e^i_a-\omega^{0i}_ae^i_0\label{torsion00a no TG}\\
&=&-\partial_aN+\omega^{0i}_0e^i_a-\omega^{0i}_aN^be^i_b,\label{torsion00a}\\
T^i_{0a}&=&\partial_0e^i_a-\partial_ae^i_0+\omega^{ij}_0e^j_a-\omega^{ij}_ae^j_0+\omega^{i0}_0e^0_a-\omega^{i0}_ae^0_0\\
&=&\partial_0e^i_a-e^i_b\partial_aN^b-N^b\partial_ae^i_b+\eps^i_{~jk}(\omega^k_0e^j_a-\omega^k_aN^be^j_b)+\omega^{0i}_aN,\label{torsioni0a}
\ea
\end{subequations}
where we have introduced the notation
\be
\omega_\mu^i\equiv\frac{1}{2}\eps^i_{~jk}\omega_\mu^{jk}.
\ee
We are going to successively recover these equations at the Hamiltonian level, in the form of solutions to second class constraints, conditions on Lagrange multipliers, and dynamical evolution equations.

The $3+1$ decomposition of the Holst action \eqref{holst action} leads to the expression
\ba\label{holst decomposed1}
S&=&\frac{1}{2}\int_\mathbb{R}\de t\int_\Sigma\de^3x\,\eps^{abc}\bigg[\left(\eps_{ijk}e_a^ie_b^jF_{0c}^{0k}+\gamma^{-1}e^i_ae_b^jF_{0c}^{ij}\right)
+N\left(\frac{1}{2}\eps_{ijk}e_a^iF_{bc}^{jk}+\gamma^{-1}e_a^iF_{bc}^{0i}\right)\nonumber\\
&&\qquad\qquad\qquad\qquad\ +N^i \left( \eps_{ijk}e_a^jF_{bc}^{0k}+\gamma^{-1}e_a^jF_{bc}^{ij}\right)\bigg].
\ea
The various components of the curvature tensor appearing in \eqref{holst decomposed1} can be written as
\begin{subequations}\label{curvatures}
\ba
 F_{\mu \nu}^{0i}&=&\partial_\mu\omega_\nu^{0i}-\partial_\nu\omega_\mu^{0i}+\eps^i_{~jk}(\omega_\nu^j\omega_\mu^{0k}-\omega_\mu^j\omega_\nu^{0k}),\\
F_{\mu \nu}^{ij}&=&\eps^{ij}_{~~k}(\partial_\mu\omega_\nu^k-\partial_\nu\omega_\mu^k)-\omega_\mu^{0i}\omega_\nu^{0j}+\omega_\nu^{0i}\omega_\mu^{0j}+\omega_\mu^j\omega_\nu^i-\omega_\nu^j\omega_\mu^i.
\ea
\end{subequations}
Naturally, it is the first term between parenthesis in \eqref{holst decomposed1} that indicates what the conjugated variables are since it contains the time derivatives. In particular, its form suggests that we introduce the $\su(2)$-valued one-form
\be\label{ABtilde}
\tilde{A}^i_\mu\equiv\omega^{0i}_\mu+\gamma^{-1}\omega^i_\mu,
\ee
as well as the densitized triad field (also known as the electric field)
\be
E^a_i\equiv\sqrt{\det(E)}\,e^a_i=\det(e)\,e^a_i=\frac{1}{2}\eps^{abc}\eps_{ijk}e_b^je_c^k.
\ee
At this point, it is useful to recall some relations that we will use later on:
\be
e^i_a=\f{\eps_{abc}\eps^{ijk}E^b_jE^c_k}{2\sqrt{\det(E)}},\qquad\eps^{abc}e^i_a=\f{\eps^{ijk}E^b_jE^c_k}{\sqrt{\det(E)}},\qquad\f{\partial_a\det(e)}{\det(e)}=e^b_i\partial_ae^i_b,\qquad\eps_{ijk}e^i_ae^j_b=\det(e)\,\eps_{abc}e^c_k.
\ee
Now, the idea is to use \eqref{ABtilde} in order to express all the connection components in \eqref{curvatures} solely in terms of $\tilde{A}^i_\mu$ and $\omega^i_\mu$. After some calculations, we arrive at the following  expression for the action:
\ba\label{action decomposed}
S
&=&\int_\mathbb{R}\de t\int_\Sigma\de^3x\bigg[E^a_i\partial_0\tilde{A}^i_a+\tilde{\alpha}^i(\partial_aE^a_i-\eps_{ij}^{~~k}\omega^j_aE^a_k)+\tilde{\beta}^i(\partial_aE^a_i-\gamma\eps_{ij}^{~~k}\tilde{A}^j_aE^a_k) \\
&&\qquad\qquad\qquad+N\frac{\eps^{imn}E^a_mE^b_n}{2\sqrt{\det(E)}}\left(2\gamma^{-1}\partial_a\tilde{A}^i_b-\eps_{ijk}\tilde{A}^j_a\tilde{A}^k_b+(1-\gamma^{-2})(2\partial_a\omega^i_b-\eps_{ijk}\omega^j_a\omega^k_b)\right)\nonumber\\
&&\qquad\qquad\qquad+N^aE^b_i\left(\partial_b\tilde{A}^i_a-\partial_a\tilde{A}^i_b+\gamma\eps^i_{~jk}\tilde{A}^j_a\tilde{A}^k_b+(\gamma^{-3}-\gamma^{-1})\eps^i_{~jk}(\omega^j_a-\gamma\tilde{A}^j_a)(\omega^k_b-\gamma\tilde{A}^k_b)\right)\bigg],\nonumber
\ea
where we have introduced the multipliers
\be\label{multipliers alphabeta}
\begin{tabular}{lllllll}
$\tilde{\alpha}^i$&$\equiv$&$(1-\gamma^{-2})(\tilde{A}^i_0-\gamma^{-1}\omega^i_0)$ 
&~~~~~~~~~&
$\tilde{\beta}^i$&$\equiv$&$\tilde{A}^i_0-\tilde{\alpha}^i$ \cr
& $=$  & $(1-\gamma^{-2})\omega^{0i}_0$,
&~~~~~~~~~&
& $=$ & $\gamma^{-2}\tilde{A}^i_0+(\gamma^{-1}-\gamma^{-3})\omega^i_0$
\cr
&&
&~~~~~~~~~&
& $=$ & $\gamma^{-2}\omega^{0i}_0+\gamma^{-1}\omega^i_0$.
\end{tabular}
\ee
The non-reduced phase space is defined by the Poisson brackets
\be
\lb E^a_i,\tilde{A}_b^j\rb=\delta^a_b\delta_i^j,\qquad\qquad\lb\pi^a_i,\omega_b^j\rb=\delta^a_b\delta_i^j,
\ee
where we have added a momentum $\pi^a_i$ conjugated to $\omega^i_a$, together with a multiplier $\mu^i_a$ in order to impose that this new momentum be vanishing. The variables $\mu^i_a$, $\tilde{\alpha}^i$, $\tilde{\beta}^i$, $N$ and $N^a$ are non-dynamical, and appear therefore as Lagrange multipliers enforcing respectively the following primary constraints\footnote{Note that we use the notation $\omega_a$ only to denote $\omega^i_a$, and never $\omega^{0i}_a$.}:
\begin{subequations}\label{holst constraint set}
\ba
\pi^a&\approx&0,\\
S&\equiv&\partial_aE^a-\omega_a\time E^a\approx0,\label{secondclassS}\\
G&\equiv&\partial_aE^a-\gamma\tilde{A}_a\time E^a\approx0,\\
C&\equiv&
\frac{E^a\time E^b}{2\sqrt{\det(E)}}\cdot\Big(2\gamma^{-1}\partial_a\tilde{A}_b-\tilde{A}_a\time\tilde{A}_b+(1-\gamma^{-2})(2\partial_a\omega_b-\omega_a\time\omega_b)\Big)\approx0,\\
\widetilde{H}_a&\equiv&
E^b\cdot\Big(\partial_b\tilde{A}_a-\partial_a\tilde{A}_b+\gamma\tilde{A}_a\time\tilde{A}_b+(\gamma^{-3}-\gamma^{-1})(\omega_a-\gamma\tilde{A}_a)\time(\omega_b-\gamma\tilde{A}_b)\Big)\approx0.~~~~~~
\ea
\end{subequations}
Before going further, let us simplify the expression of these primary constraints. We see from the constraints $G$ that in order to obtain the Gauss law, the connection that we need to choose is the variable $A\equiv-\gamma\tilde{A}$. In term of this new variable, the constraints $G$, $C$ and $\widetilde{H}_a$ become respectively
\begin{subequations}
\ba
G&=&D_aE^a\equiv\partial_aE^a+A_a\time E^a,\\
C&=&-\frac{\gamma^{-2}E^a\time E^b}{2\sqrt{\det(E)}}\cdot\left(\mathcal{F}_{ab}+(1-\gamma^2)R_{ab}\right),\label{scalar constraint AB}\\
\widetilde{H}_a&=&\gamma^{-1}E^b\cdot\mathcal{F}_{ab}+(\gamma^{-3}-\gamma^{-1})(G-S)\cdot(\omega_a+A_a),
\ea
\end{subequations}
where
\be
\mathcal{F}_{ab}^i\equiv\partial_aA^i_b-\partial_bA^i_a+\eps^i_{~jk}A_a^jA_b^k,\qquad\qquad
R_{ab}^i\equiv\partial_a\omega_b^i-\partial_b\omega^i_a-\eps^i_{~jk}\omega^j_a\omega_b^k.
\ee
The notation $R_{ab}^i$ as well as the sign difference between these two expressions will become obvious right below, when we solve the second class constraints. For the time being, we can write the total Hamiltonian as
\be\label{Htot}
-H_\text{tot}\equiv\int_\Sigma\de^3x\left(\mu^i_a\pi^a_i+\alpha^iS_i+\beta^iG_i+NC+N^aH_a\right),
\ee
where the new Lagrange multipliers $\alpha^i$ and $\beta^i$ have been defined as
\be\label{multipliers alphabeta}
\begin{tabular}{lllllll}
$\alpha^i$&$\equiv$&$\tilde{\alpha}^i-(\gamma^{-3}-\gamma^{-1})N^a(\omega^i_a+A^i_a)$ 
&~~~~~~~~~&
$\beta^i$&$\equiv$&$\tilde{\beta}^i+(\gamma^{-3}-\gamma^{-1})N^a(\omega^i_a+A^i_a)$ \cr
& $=$  & $(1-\gamma^{-2})(\omega^{0i}_0-N^a\omega^{0i}_a)$,
&~~~~~~~~~&
& $=$ & $\gamma^{-2}\omega^{0i}_0+\gamma^{-1}\omega^i_0+(1-\gamma^{-2})N^a\omega^{0i}_a$ \cr
&& 
&~~~~~~~~~&
& $=$ & $-\gamma^{-1}A^i_0+(1-\gamma^{-2})(N^a\omega^{0i}_a-\omega^{0i}_0)$,
\end{tabular}
\ee
and the new version $H_a$ of the vector constraint is $H_a\equiv\gamma^{-1}E^b\cdot\mathcal{F}_{ab}$.

\subsubsection{Vanishing of $T^i_{ab}$}
\label{subsubsec:second class}

\noindent Let us look at the time evolution of the primary constraints. In fact, one can show that the only constraints that lead to secondary constraints are $\pi^a_i\approx0$. The secondary constraints are obtained by computing the time evolution $\partial_0\pi^a_i=\lb H_\text{tot},\pi^a_i\rb$, which is given by
\ba\label{secondary}
\partial_0\pi^a_i&=&-\int_\Sigma\de^3x\left(\lb\alpha^jS_j,\pi^a_i\rb+\lb NC,\pi^a_i\rb\right)\nonumber\\
&=&-\int_\Sigma\de^3x\left(\alpha^j\lb-\eps_{jk}^{~~l}\omega^k_bE^b_l,\pi^a_i\rb+\frac{1}{2}(1-\gamma^{-2})\lb N\eps^{bcd}e^j_dR^j_{bc},\pi^a_i\rb\right)\nonumber\\
&=&\eps_{ij}^{~~k}\alpha^jE^a_k+(1-\gamma^{-2})\eps^{abc}\left(N(\partial_be^i_c+\eps^i_{~jk}e^j_c\omega^k_b)+e^i_c\partial_bN\right).
\ea
Multiplying this expression by $E^e_i$ and symmetrizing the spatial indices, we obtain the 6 secondary constraints
\be\label{Psiconstraints}
\Psi^{ab}\equiv\eps^{(acd}E^{b)}_i(\partial_ce^i_d+\eps^i_{~jk}e^j_d\omega^k_c)=\f{1}{\sqrt{\det(E)}}\eps^{ijk}E^{(b}_i\left(\partial_cE^{a)}_j-\eps_{jl}^{~~m}\omega^l_cE^{a)}_m\right)E^c_k\approx0.
\ee
These second class constraints, together with $\eqref{secondclassS}$, imply that the rotational part of the spatial gauge connection is the metric compatible Levi-Civita connection, i.e.
\ba\label{levi-civita solution}
\omega^i_a&=&-\Gamma^i_a\nonumber\\
&=&-\f{1}{2}\eps^{ijk}e^b_k(\partial_be^j_a-\partial_ae^j_b+e^c_je^l_a\partial_be^l_c)\\
&=&-\f{1}{2}\eps^{ijk}E^b_k(\partial_bE^j_a-\partial_aE^j_b+E^c_jE^l_a\partial_bE^l_c)
-\f{1}{4}\eps^{ijk}E^b_k\left(2E^j_a\f{\partial_b\det(E)}{\det(E)}-E^j_b\f{\partial_a\det(E)}{\det(E)}\right).\nonumber
\ea
With the resolution of these second class constraints (which are conjugated to the 9 second class constraints $\pi^a_i\approx0$), we recover the vanishing of the torsion components $T^i_{ab}$ given by \eqref{torsioniab}. Note that it is possible to avoid solving directly the second class constraints by computing instead the associated Dirac bracket. This might be an interesting way to make contact with the covariant loop quantization developped by Alexandrov. We compute the Dirac bracket for these second class constraints in appendix \ref{appendix:Dirac}.

\subsubsection{Vanishing of $T^0_{ab}$}

\noindent The torsion equation on the components $T^0_{ab}$ can now be obtained by combining the Gauss constraint $G$ with the second class constraint $S$. By subtracting these two constraints, we obtain the condition $\eps^{ij}_{~~k}\omega^{0j}_aE^a_k=0$. Assuming that the triad is invertible, this expression leads to $\eps^{abc}\omega^{0i}_ae^i_b=0$, which is indeed equivalent to \eqref{torsion0ab}.

\subsubsection{Vanishing of $T^0_{0a}$}

\noindent Now, there are still 3 equations to be extracted from the requirement that the primary constraint $\pi^a_i\approx0$ be preserved in time, and they can be recovered by multiplying \eqref{secondary} by $E^e_i$ without symmetrizing the indices. This gives
\be
-\alpha^ie^i_a+(1-\gamma^{-2})\partial_aN=0.
\ee
These are equations fixing some of the Lagrange multipliers, and using \eqref{multipliers alphabeta} it is easy to see that they imply
\be\label{multipliers fixed}
-\partial_aN+\omega^{0i}_0e^i_a-\omega_b^{0i}N^be^i_a=0.
\ee
Using now the vanishing of the components \eqref{torsion0ab}, we can write $\omega^{0i}_ae^i_b=\omega^{0i}_be^i_a$, and then \eqref{multipliers fixed} finally yields the vanishing of the torsion components $T^0_{0a}$.

\subsubsection{Vanishing of $T^i_{0a}$}
\label{subsec torsion dyn}

\noindent The last components of the torsion are hidden in the dynamics of the theory. In order to see how this comes about, let us look at the dynamical evolution of the triad $e^i_a$, which is generated by commutation with the total Hamiltonian. For this, we need to know the following Poisson brackets:
\be
\lb e^i_a,A^j_b\rb=\f{\gamma}{2\det(e)}(2e^i_be^j_a-e_a^ie_b^j).
\ee
Before computing the evolution of the triad under the various constraints entering the Hamiltonian \eqref{Htot}, it is useful to first derive the general expression
\be
\int_\Sigma\de^3x\,\lb n_j\mathcal{F}^j_{ab},e^i_c\rb=\f{\gamma}{\det(e)}D_{[a}n_j(2e^i_{b]}e^j_c-e^j_{b]}e^i_c),
\ee
which is valid for any smearing function $n(x)$ (possibly carrying spatial indices) that does not depend on the connection $A$. Introducing the smeared constraints
\be
G(u)\equiv\int_\Sigma\de^3x\,u^iG_i,\qquad\qquad C(N)\equiv\int_\Sigma\de^3x\,NC,\qquad\qquad H(\vec{N})\equiv\int_\Sigma\de^3x\,N^aH_a,
\ee
we can then compute
\begin{subequations}
\ba
\lb G(u),e^i_a\rb&=&-\gamma\eps^i_{~jk}u^je^k_a,\\
\lb C(N),e^i_a\rb&=&\omega^{0i}_aN+\gamma^{-1}\eps^{ij}_{~~k}e^b_je^k_a\partial_bN,\\
\lb H(\vec{N}),e^i_a\rb&=&-e^i_b\partial_aN^b-N^bD_be^i_a=-e^i_b\partial_aN^b+\gamma N^b\eps^i_{~jk}\omega^{0j}_be^k_a,
\ea
\end{subequations}
where in the last equation we have used the solution \eqref{levi-civita solution} to the second class constraints to simplify the covariant derivative of the triad. The dynamical evolution of $e^i_a$ can now be computed, and, using the expression \eqref{multipliers alphabeta} for the multipliers $\beta^i$, we find that\footnote{Note that there is no contribution from $G_j\lb\beta^j,e^i_a\rb$ since this latter can be shown to be vanishing upon use of the constraints \eqref{secondclassS} and the condition \eqref{torsion0ab}.}
\ba
\partial_0e^i_a&\equiv&\lb H_\text{tot},e_a^i\rb\nonumber\\
&\equiv&-\lb G(\beta),e_a^i\rb-\lb C(N),e_a^i\rb-\lb H(\vec{N}),e_a^i\rb\nonumber\\
&=&\eps^i_{~jk}(\gamma^{-1}\omega^{0j}_0+\omega^j_0-\gamma^{-1}N^b\omega^{0j}_b)e^k_a-\omega^{0i}_aN-\gamma^{-1}\eps^{ij}_{~~k}e^b_je^k_a\partial_bN+e^i_b\partial_aN^b.\label{dynamical torsion components}
\ea
Using the condition \eqref{multipliers fixed} on the multipliers to express $\partial_bN$, we find that this equation is equivalent to the vanishing of the last components \eqref{torsioni0a} of the torsion, i.e. $T^i_{0a}=0$.

Finally, for the sake of completeness, let us finish this subsection by rewriting the scalar constraint in the form that usually appears in loop quantum gravity, i.e.
\be
C=-\frac{E^a\time E^b}{2\sqrt{\det(E)}}\cdot\left(\mathcal{F}_{ab}-(1-\gamma^2)K_a\time K_b\right)+(1-\gamma^{-2})\partial_a\left(\f{E^a}{\sqrt{\det(E)}}\right)\cdot G,
\ee
where $K^i_a\equiv\omega^{0i}_a$.

The meaning of this computation is that, from the classical point of view, if we perform the canonical analysis properly we recover all the torsion-free equations that lead to the elimination of the Barbero-Immirzi parameter. At the quantum level, the story is different because we do not have access to the equations involving the Lagrange multipliers. Therefore, we can wonder what happens if we relax the vanishing of the components $T^0_{0a}$ of the torsion, which are the ones left aside if we do not consider the fixation of the Lagrange multipliers in the Hamiltonian analysis. In this case, one can see from \eqref{torsion components} that the components $\omega_0^{0i}$ of the connection cannot be written in terms of the tetrad. Moreover, the form \eqref{Htot} of the Hamiltonian shows that these components are involved in the expression
\be
(\gamma-\gamma^{-1})\omega_0^{0i}(S_i-G_i)=(\gamma-\gamma^{-1})\omega_0^{0i}\eps_{ij}^{~~k}\omega^{0j}_aE^a_k,
\ee
which contains both first class and second class constraints. If these first and second class constraints are resolved at the classical level, the dependency on $\omega_0^{0i}$ drops out of the theory, and the non-imposition of $T^0_{0a}\approx0$ is not problematic. However, in the quantum theory, where only the second class constraints $S_i$ are imposed strongly (already at the classical level), the Gauss constraint is allowed to fluctuate, and the undetermined components $\omega_0^{0i}$ will not disappear.

\subsection{Gauge symmetries and constraint algebra}

\noindent Now that we have analyzed the second class constraints and discussed their resolution, let us discuss the first class constraints and the associated symmetries. For this purpose, we start by recalling the algebra of (first class) constraints. It is given by \cite{AL}
\begin{subequations}
\ba
\lb G(u),G(v)\rb&=&\gamma G(u\time v),\\
\lb H_\text{diff}(\vec{N}),G(u)\rb&=&G(\pounds_{\vec{N}}u),\\
\lb\widetilde{C}(N),G(u)\rb&=&0,\\
\lb H_\text{diff}(\vec{N}),H_\text{diff}(\vec{M})\rb&=&H_\text{diff}(\pounds_{\vec{N}}\vec{M}),\\
\lb\widetilde{C}(N),H_\text{diff}(\vec{N})\rb&=&\widetilde{C}(\pounds_{\vec{N}}N),\\
\lb\widetilde{C}(N),\widetilde{C}(M)\rb&=&H_\text{diff}(\vec{U})+G(U^aA_a)+(\gamma^{-2}-1)G\left(\f{E^a\partial_aN\time E^b\partial_bM}{\det(E)}\right),
\ea
\end{subequations}
where
\be
U^a\equiv\f{E^a_iE^b_i}{\sqrt{\det(E)}}(N\partial_bM-M\partial_bN).
\ee
Note that for the scalar constraint we have used the usual expression
\be\label{new4d scalar}
\widetilde{C}\equiv-\frac{E^a\time E^b}{2\sqrt{\det(E)}}\cdot\left(\mathcal{F}_{ab}-(1-\gamma^2)K_a\time K_b\right),
\ee
which differs from $C$ by a term proportional to the Gauss constraint. 

The algebra, although it is not a Lie algebra because it has field dependent structure functions, is obviously closed. As usual, the constraints generate the symmetries of the theory. The generator of internal $\SU(2)$ gauge transformations is the Gauss constraint. It acts on the connection and the densitized triad as
\be\label{action gauss}
\lb G(u),A^i_a\rb=\gamma D_au^i,\qquad\qquad\lb G(u),E^a_i\rb=-\gamma\eps_{ij}^{~~k}u^jE_k^a.
\ee
The generator of spatial diffeomorphisms is given by the constraint
\be
H_\text{diff}(\vec{N})\equiv\int_\Sigma\de^3x\,N^a(H_a-\gamma^{-1}A^i_aG_i),
\ee
and one can check that we indeed have
\begin{subequations}
\ba
\lb H_\text{diff}(\vec{N}),A_a^i\rb&=&-N^b\partial_bA_a^i-A_b^i\partial_aN^b=-\pounds_{\vec{N}}A_a^i,\label{action diffA}\\
\lb H_\text{diff}(\vec{N}),E^a_i\rb&=&-N^b\partial_bE^a_i-E^a_i\partial_bN^b+E^b_i\partial_bN^a=-\pounds_{\vec{N}}E^a_i.
\ea
\end{subequations}
Now we still have to discuss the time diffeomorphisms that are generated by the total Hamiltonian. It is known since the work of Samuel \cite{samuel}, and later on the work of Alexandrov in the Lorentz-covariant formulation \cite{alexandrov5}, that the Ashtekar-Barbero is not a spacetime connection, in the sense that it does not transform as a one-form under the action of the generator of time diffeomorphisms. In the next subsection we re-derive this result in the time gauge.

\subsection{Transformation properties of the connection}

\noindent Let us first consider the self-dual case $\gamma=1$ (in the physically relevant case of Lorentzian signature, the (anti) self-dual value is $\gamma=\pm i$), and introduce the smeared total Hamiltonian
\ba
-H_\text{tot}(\xi)&=&\int_\Sigma\de^3x\,\xi\left(\beta^iG_i+NC_\text{\tiny{SD}}+N^aH_a\right)\nonumber\\
&=&\int_\Sigma\de^3x\,\xi\left((\beta^i+N^aA^i_a)G_i+NC_\text{\tiny{SD}}+N^a(H_a-A^i_aG_i)\right)\nonumber\\
&=&G\big(\xi(\beta+N^aA_a)\big)+C_\text{\tiny{SD}}(\xi N)+H_\text{diff}(\xi\vec{N})\nonumber\\
&=&G\big(\xi(-A_0+N^aA_a)\big)+C_\text{\tiny{SD}}(\xi N)+H_\text{diff}(\xi\vec{N}),
\ea
where $C_\text{\tiny{SD}}$ is the self-dual part of the scalar constraint, i.e. the constraint obtained by setting $\gamma=1$ in \eqref{scalar constraint AB}. Using the transformation laws \eqref{action gauss} and \eqref{action diffA}, one can show that\footnote{Notice that $\lb C_\text{\tiny{SD}}(\xi N),A^i_a\rb=\xi\lb C_\text{\tiny{SD}}(N),A^i_a\rb$ because $C_\text{\tiny{SD}}$ does not contain derivatives of $E^a_i$.}
\be
\lb H_\text{tot}(\xi),A^i_a\rb=A^i_0\partial_a\xi+\xi\left(D_aA^i_0-\lb C_\text{\tiny{SD}}(N),A^i_a\rb+N^b\partial_bA^i_a-N^bD_aA^i_b\right).
\ee
However, we know from the computation \eqref{dynamical torsion components} of subsection \ref{subsec torsion dyn} that $H_\text{tot}(1)$ is the generator of time evolution. For this reason, we have that $\lb H_\text{tot}(1),A^i_a\rb=\partial_0A^i_a$. Furthermore, an explicit computation shows that this Poisson bracket is given by
\be
\lb H_\text{tot}(1),A^i_a\rb=D_aA^i_0-\lb C_\text{\tiny{SD}}(N),A^i_a\rb+N^b\partial_bA^i_a-N^bD_aA^i_b.
\ee
We can therefore conclude that $\lb H_\text{tot}(\xi),A^i_a\rb=A^i_0\partial_a\xi+\xi\partial_0A^i_a$, as expected for a spacetime connection.

To study the case $\gamma\neq1$, it is useful to write the scalar constraint as $C=\gamma^{-2}C_\text{\tiny{SD}}+C_\gamma$. Because the part $C_\gamma$ contains derivatives of $E^a_i$ (once $\omega^{0i}_a$ is replaced by $\gamma^{-1}(\Gamma^i_a(E)-A^i_a)$), the Poisson bracket $\lb C(\xi N),A^i_a\rb$ will necessarily be of the form 
\be
\gamma^{-2}\lb C_\text{\tiny{SD}}(\xi N),A^i_a\rb+(1-\gamma^2)\big(\xi Nf^i_a+\xi g^i\partial_aN+Ng^i\partial_a\xi\big),
\ee
where $f^i_a$ and $g_i$ are certain functions of the electric field (whose expressions are not important for our discussion). Now, using the total Hamiltonian
\be
-H_\text{tot}(\xi)=G\big(\xi(\beta+\gamma^{-1}N^aA_a)\big)+C(\xi N)+H_\text{diff}(\xi\vec{N}),
\ee
we can compute
\ba
\delta_\xi A^i_a&\equiv&\lb H_\text{tot}(\xi),A^i_a\rb\nonumber\\
&=&A^i_0\partial_a\xi+\xi\left(-\gamma D_a\beta^i-\gamma^{-2}\lb C_\text{\tiny{SD}}(N),A^i_a\rb+N^b\partial_bA^i_a-N^bD_aA^i_b-(1-\gamma^2)(Nf^i_a+g^i\partial_aN)\right)\nonumber\\
&&+\Big((\gamma^{-1}-\gamma)(N^a\omega^{0i}_a-\omega^{0i}_0)-(1-\gamma^2)Ng^i\Big)\partial_a\xi.
\ea
Similarly to what happens in the self-dual case, the explicit expression for $\lb H_\text{tot}(1),A^i_a\rb=\partial_0A^i_a$ turns out to be given by the term proportional to $\xi$ in the above equation. We therefore obtain that
\be
\lb H_\text{tot}(\xi),A^i_a\rb=A^i_0\partial_a\xi+\xi\partial_0A^i_a+(1-\gamma^2)\Big(\gamma^{-1}(N^a\omega^{0i}_a-\omega^{0i}_0)-Ng^i\Big)\partial_a\xi,
\ee
which shows that $A^i_a$ does not transform as a spacetime connection. This expression should be compared with its Lorentz-covariant counterpart, which is equation (48) of \cite{alexandrov5}.

\subsection{Discussion}

\noindent As mentioned in the introduction, the status of the Barbero-Immirzi parameter in loop quantum gravity is very intriguing. At the classical level, $\gamma$ is totally irrelevant, and it disappears from the theory when one goes from the first order formulation to the second order formulation of gravity. This is manifest in both the Lagrangian and the Hamiltonian framework. Here, we showed in particular how to recover all the components of the torsion-free equations in the canonical approach. Even if the components $T^i_{ab}\approx0$ are strongly imposed (at the classical level) because they appear as second class constraints, the remaining components are forced to vanish by the dynamics, the Gauss constraint, and the fixation of Lagrange multipliers. As a consequence, these components are never strongly imposed at the classical level (which is absolutely normal), and therefore we can expect their imposition in the quantum theory not to totally eliminate the $\gamma$-dependency due to the ``quantum fluctuations''.

Moreover, it appears that there are 3 components of the torsion-free equations ($T^0_{0a}\approx0$) that can never be imposed at the quantum level when we work in the time gauge. These are equations fixing some of the Lagrange multipliers, and as such they can be derived neither from the (first or second class) constraints, nor from the dynamics. This is in fact a consequence of working in the time gauge. In the Lorentz-covariant approach to loop quantum gravity, these 3 components are fixed by the  dynamics, and more precisely by the time evolution of the variable $\chi$ of \eqref{tetrad decomposition} (indeed, notice that $\partial_0e^0_a$ appears in \eqref{torsion00a no TG}, but not in \eqref{torsion00a}). We do not know if this can generate anomalies in the quantum theory. In any case, as discussed in the previous paragraph, this should not influence the $\gamma$-dependency of the quantum theory at the dynamical level. This issue should however still be investigated more deeply, since it is rather intriguing that it appears only in the time gauge, and not in the Lorentz-covariant formulation.

Another striking difference between the time gauge formulation and the Lorentz-covariant formulation of the Holst action concerns the transformation properties of the connection. This is in fact a well-known observation. The Ashtekar-Barbero connection, wether it is written in the time gauge or in the covariant formalism, does not transform properly under time diffeomorphisms. Here we have derived this result in the time gauge. This lack of proper transformation behavior of the connection could be a serious problem at the level of the dynamics, and regardless of the imposition of the above-mentioned torsion-free equations, it could lead to anomalies in the quantum theory. What is clear is that the only known way to construct a connection (different from the original spin connection or from the self-dual connection) that transforms properly under time diffeomorphisms, is to relax the time gauge condition. Unfortunately, nothing is known about the quantization of the canonical theory defined with the shifted connection of Alexandrov.

To close this discussion, let us stress that there seem to be a very subtle interplay between the Barbero-Immirzi parameter $\gamma$, the time gauge, and the (classical and quantum) dynamics. However, four-dimensional gravity is obviously too complicated to clarify this. For this reason, in the following section we propose a model of gravity in three dimensions where there is a Barbero-Immirzi parameter that plays the same role as in the four-dimensional case. Because three-dimensional gravity is a (non-trivial) totally integrable system, there is hope that we can learn more about the fate of $\gamma$ in this framework.

\section{The Barbero-Immirzi ambiguity in three-dimensional gravity}
\label{sec:2}

\noindent As we just said above, in order to investigate the issues related to the Barbero-Immirzi parameter and the choice of the time gauge, let us introduce the three-dimensional action
\be\label{3d action}
S=\int_\mathcal{M}\de^3x\,\eps^{\mu\nu\rho}\left(\f{1}{2}\eps_{IJKL}x^Ie^J_\mu F^{KL}_{\nu\rho}+\gamma^{-1}x^Ie^J_\mu F^{IJ}_{\nu\rho}\right),
\ee
where $x^I\in\mathbb{R}^4$. In \cite{GN}, it was shown that this action is equivalent to three-dimensional Euclidean gravity. In appendix \ref{appendix symmetry red}, we show that it can also be obtained from a symmetry reduction of the four-dimensional Holst action.

In this section, we would like to illustrate the interplay between the choice of internal gauge and the role of the Barbero-Immirzi parameter at the classical level. For this, we consider two different gauges, and show that they lead to two rather different descriptions of the (same) phase space. In the first case, the parameter $\gamma$ disappears completely from the phase space, whereas with the second gauge choice it appears exactly in the same way as it does in the four-dimensional theory. Thus, this model provides us with a description of the classical phase space of three-dimensional gravity with a non-trivial $\gamma$-dependency.

Before going further, let us point out that the action \eqref{3d action} is invariant under the following symmetries:
\begin{itemize}
\item[$\bullet$] A rescaling symmetry, generated by a scalar $\alpha$ and acting like
\be\label{symm3d1}
e^I_\mu\longrightarrow\alpha e^I_\mu,\qquad\qquad x^I\longrightarrow\f{1}{\alpha}x^I.
\ee
\item[$\bullet$] Three translational symmetries, generated by a vector $\beta_\mu$ and acting like
\be\label{symm3d2}
e^I_\mu\longrightarrow e^I_\mu+\beta_\mu x^I.
\ee
\end{itemize}
These Lagrangian symmetries can be identified at the Hamiltonian level if one performs the canonical analysis of the action \eqref{3d action} and treats $x^I$ as a dynamical variable. This analysis being quite involved, we choose not to reproduce it here. One should however keep in mind that the symmetries \eqref{symm3d1} and \eqref{symm3d2} are present, because they will be used in subsection \ref{subsec:gamma-gauge} in order to define the $\gamma$-dependent gauge.

\subsection{$\boldsymbol{\gamma}$-independent gauge}

\noindent Let us consider the gauge $x^I=(1,0,0,0)$. With this choice, the internal $\SO(4)$ symmetry group is broken into an $\SU(2)$ subgroup (the one that stabilizes $x^I$), and the action reduces to
\be\label{red1}
S=\int_\mathcal{M}\de^3x\,\eps^{\mu\nu\rho}\left(\f{1}{2}\eps_{ijk}e^i_\mu F^{jk}_{\nu\rho}+\gamma^{-1}e^i_\mu F^{0i}_{\nu\rho}\right),
\ee
where the curvature components are defined as in \eqref{curvatures}. Introducing $E^a_i\equiv2\eps^{ab}e^i_b$ and $A^i_a\equiv\omega^i_a+\gamma^{-1}\omega^{0i}_a$, and using the expression
\be
F_{\mu\nu}^{ij}=\eps^{ij}_{~~k}\left(\partial_\mu\omega_\nu^k-\partial_\nu\omega_\mu^k+\eps^k_{~lm}\omega_\nu^l\omega_\mu^m-\eps^k_{~lm}\omega_\mu^{0l}\omega_\nu^{0m}\right),
\ee
a straightforward calculation shows that the $2+1$ action can be written in the following three-dimensional Ashtekar-Barbero form:
\ba
S&=&\int_\mathcal{M}\de^3x\,\eps^{\mu\nu\rho}e_\mu^i\left(\partial_\nu A^i_\rho-\partial_\rho A^i_\nu-\eps^i_{~jk}A_\nu^jA_\rho^k+(\gamma^{-2}-1) \eps^i_{~jk}\omega_\nu^{0j}\omega_\rho^{0k}\right)\nonumber\\
&=&\int_\mathbb{R}\de t\int_\Sigma\de^2x\left(E^a_i\partial_0A^i_a+A^i_0G_i+\omega^{0i}_0S_i+\eps^{ab}e^i_0\left[\mathcal{F}^i_{ab}+(\gamma^{-2}-1)\eps^i_{~jk}\omega^{0j}_a\omega^{0k}_b\right]\right),\label{3dAB action}
\ea
where
\begin{subequations}
\ba
G&\equiv&\partial_aE^a-A_a\time E^a,\\
\mathcal{F}_{ab}&\equiv&\partial_aA_b-\partial_bA_a-A_a\time A_b,\\
S_i&\equiv&(\gamma^{-2}-1)\eps_{ij}^{~~k}\omega^{0j}_aE^a_k.
\ea
\end{subequations}
In the action \eqref{3dAB action}, the multipliers $A^i_0$ are enforcing the $\SU(2)$ Gauss constraint $G_i$. Similarly, the multipliers $e^i_0$ are enforcing a scalar constraint analogous to the four-dimensional constraint \eqref{new4d scalar}. We see that for the (anti) self-dual value $\gamma=\pm1$, the action \eqref{3dAB action} reduces to that of $\SU(2)$ BF theory. Let us see what happens for a generic value of the Barbero-Immirzi parameter. The total Hamiltonian is given by
\be
-H_\text{tot}\equiv\int_\Sigma\de^3x\left(\mu^i_a\pi^a_i+\omega^{0i}_0S_i+A^i_0G_i+\eps^{ab}e^i_0\left[\mathcal{F}^i_{ab}+(\gamma^{-2}-1)\eps^i_{~jk}\omega^{0j}_a\omega^{0k}_b\right]\right),
\ee
where we have introduced the momenta $\pi^a_i$ conjugated to the components $\omega^{0i}_a$, together with the multipliers $\mu^i_a$. One can see that the preservation of the primary constraints $\pi^a_i\approx0$ leads to the 6 secondary constraints
\be
\eps_{ijk}(\omega^{0j}_0E^a_k+2\eps^{ab}e^j_0\omega^{0k}_b)\approx0.
\ee
From these 6 constraints, one can extract 3 conditions involving Lagrange multipliers, and the 3 secondary constraints
\be
\Psi^{ab}\equiv\eps^i_{~jk}\eps^{(ab}E^{c)}_i\omega^{0k}_be^j_0\approx0.
\ee
Together with $S_i$ and $\pi^a_i\approx0$, these constraints form a second class system, and imply in particular that the components $\omega^{0i}_a$ of the connection are vanishing. Therefore, the dependency on the Barbero-Immirzi parameter completely disappears, and we are left with three-dimensional $\su(2)$ BF theory, i.e. first order three-dimensional gravity.

Notice that the gauge that we are considering here does not affect directly the variables $e_\mu^I$. Indeed, we do not explicitly set certain components of $e_\mu^I$ to zero, as it is the case with the time gauge in four dimensions. In this sense, the gauge $x^I=\delta^I_J$ keeps a kind of covariance, which is probably the reason for which the Barbero-Immirzi parameter $\gamma$ disappears already at the kinematical level. Now, we wonder if there is another gauge choice that also leads to a Hamiltonian theory which is equivalent to three-dimensional gravity, but with an explicit dependency on the Barbero-Immirzi parameter. We show in the next subsection that this is indeed possible, and we are going to see that the gauge choice affects directly the components $e_\mu^I$.

To finish, let us stress that one obtains the same formulation of the classical phase space for any gauge of the form $x^I=\delta^I_J$, where $J$ is a fixed value in $\{0,1,2,3\}$. In fact, it works in the same way when one fixes $x^I$ to be any (non-dynamical) vector.

\subsection{$\boldsymbol{\gamma}$-dependent gauge}
\label{subsec:gamma-gauge}

\noindent In this subsection, we show that it is possible to find a gauge fixing in the action \eqref{3d action} that yields a phase space with an explicit dependency on the Barbero-Immirzi parameter. The idea consists in finding the analog of the four-dimensional time gauge in our three-dimensional model. This can be done by choosing a particular decomposition of the one-form $e^I_\mu$, in which the time components $e^I_0$ are given by
\be\label{gauge1}
e^I_0=(e^0_0,e^i_0)\equiv(0,e^i_0)=(0,e^1_0,e^2_0,N/2),
\ee
and the spatial components $e^I_a$ are
\be\label{gauge2}
e^I_a=(e^0_a,e^i_a)=(e^0_a,e^1_a,e^2_a,e^3_a)\equiv(0,e^1_a,e^2_a,0).
\ee
The condition $e^0_a=0$ that we impose here is analogous to the time gauge $\chi=0$ that is used in the four-dimensional decomposition \eqref{tetrad decomposition}. Additionally, we will choose the internal vector $x^I$ such that
\be\label{gauge3}
x^I=(x^0,x^i)=(x^0,x^1,x^2,x^3)\equiv(1,x^1,x^2,0).
\ee
The meaning of the gauge choice encoded in the previous three equations is as follows. We are ultimately interested in describing the phase space of three-dimensional gravity, on which the gauge symmetries are the spacetime diffeomorphisms and the $\SU(2)$ gauge transformations. However, the symmetries of the action \eqref{3d action} include also the boost tranformations as well as the transformations \eqref{symm3d1} and \eqref{symm3d2}. Naturally, these can be eliminated by fixing 7 conditions. In particular, the 3 boost degrees of freedom can be used to set $e^3_a=0$ and $x^3=0$, the symmetry \eqref{symm3d1} can be used to set $x^I=(1,x^1,x^2,0)$, and the symmetry \eqref{symm3d2} with $\beta_\mu=-e^0_\mu$ can be used to set $e^0_\mu=0$ in \eqref{gauge1} and \eqref{gauge2}.

With this gauge choice, a simple calculation shows that the action \eqref{3d action} can be written in the form
\be\label{3d action 2}
S=\int_\mathcal{M}\de^3x\,\eps^{\mu\nu\rho}\Big[2(x\time e_\mu+\gamma^{-1}e_\mu)\cdot(\partial_\nu K_\rho-K_\nu\time\omega_\rho)
+(e_\mu+\gamma^{-1}x\time e_\mu)\cdot(2\partial_\nu\omega_\rho-\omega_\nu\time\omega_\rho-K_\nu\time K_\rho)\Big],
\ee
where $K^i_\mu\equiv\omega^{0i}_\mu$. In this expression for the action, the canonical term (i.e. the part containing time derivatives of the canonical variables) can be written as
\be\label{canonical term 3d}
\widetilde{E}^a_i\partial_0(\omega^i_a+\gamma^{-1}K^i_a)+X^a_i\partial_0(K^i_a+\gamma^{-1}\omega^i_a),
\ee
where $\widetilde{E}^a_i\equiv2\eps^{ab}e^i_b$, and $X^a_i\equiv2\eps^{ab}(x\time e_b)_i$. Now, because of our gauge choice, one can see that the components $i=3$ are singled out, and that we have
\be
\widetilde{E}^a_i=(\widetilde{E}^a_1,\widetilde{E}^a_2,0),\qquad\qquad X^a_i=(0,0,X^a_3)=\big(0,0,2\eps^{ab}(x^1e^2_b-x^2e^1_b)\big).
\ee
Introducing the new canonical variables
\be
E^a_i\equiv(\widetilde{E}^a_1,\widetilde{E}^a_2,X^a_3),\qquad\qquad\tilde{A}^i_a\equiv(\omega^1_a+\gamma^{-1}K^1_a,\omega^2_a+\gamma^{-1}K^2_a,K^3_a+\gamma^{-1}\omega^3_a),
\ee
the canonical term \eqref{canonical term 3d} can therefore be written in the simple form $E^a_i\partial_0\tilde{A}^i_a$. Now, the constraints enforced by the multipliers $\omega_0$, $K_0$, and $e_0$ in \eqref{3d action 2} have to be appropriately rewritten in order to mimic the structure of the four-dimensional Holst theory. A rather lengthy calculation (some details are given in appendix \ref{appendix details 3d}) shows that this can be done, and that the action can be written in the form
\be\label{3d gamma action}
S=\int_\mathbb{R}\de t\int_\Sigma\de^2x\left(E^a_i\partial_0\tilde{A}^i_a+\tilde{\alpha}^iS_i+\tilde{\beta}^iG_i+NC+N^a\widetilde{H}_a\right),
\ee
where the multipliers are
\be\label{multipliers ab 3d}
\tilde{\alpha}^i\equiv(1-\gamma^{-2})(\omega^1_0,\omega^2_0,K^3_0),
\qquad\qquad
\tilde{\beta}^i\equiv(K^1_0,K^2_0,\omega^3_0)+\gamma^{-2}(\omega^1_0,\omega^2_0,K^3_0).
\ee
The phase space is therefore parametrized by the canonical pairs
\be
\lb E^a_i,\tilde{A}^j_b\rb=\delta^a_b\delta^j_i,\qquad\qquad\lb\pi^a_i,\Omega_b^j\rb=\delta^a_b\delta_i^j,
\ee
where $\Omega^i_a\equiv(K^1_a,K^2_a,\omega^3_a)$. The primary constraints are similar to that of the set \eqref{holst constraint set}, and are given by
\begin{subequations}
\ba
\pi^a&\approx&0,\\
S&\equiv&\partial_aE^a-\Omega_a\time E^a\approx0,\\
G&\equiv&\partial_aE^a-\gamma\tilde{A}_a\time E^a\approx0,\\
C&\equiv&\frac{E^1\time E^2}{\det(\widetilde{E})}\cdot\Big(2\gamma^{-1}\partial_1\tilde{A}_2-\tilde{A}_1\time\tilde{A}_2+(1-\gamma^{-2})(2\partial_1\Omega_2-\Omega_1\time\Omega_2)\Big)\approx0,\\
\widetilde{H}_a&\equiv&E^a\cdot\Big(\partial_1\tilde{A}_2-\partial_2\tilde{A}_1-\gamma\tilde{A}_1\time\tilde{A}_2+(\gamma^{-1}-\gamma^{-3})(\Omega_1-\gamma\tilde{A}_1)\time(\Omega_2-\gamma\tilde{A}_2)\Big)\approx0.~~~~~~
\ea
\end{subequations}
To obtain the constraints $C$ and $\widetilde{H}_a$, we have used the decomposition $2e^i_0H^i_{12}=N^a\widetilde{H}_a+NC$ given in \eqref{decomposition H final}.

Now, just like in the four-dimensional case, the only primary constraints that generate secondary constraints are $\pi^a_i\approx0$. Computing their time evolution leads to the 3 new secondary constraints
\be
\partial_0\pi^{(a}\cdot E^{b)}\approx\partial_c\big(E^{(a}\time E^c\big)\cdot E^{b)}-\Omega_c\time\big(E^{(a}\time E^c\big)\cdot E^{b)}\approx0.
\ee
These 3 constraints can be combined with the 3 second class constraints $S_i\approx0$ to explicitly resolve $\Omega^i_a$ in terms of $E^a_i$. This implies in fact that $\Omega^i_a$ is the two-dimensional Levi-Civita connection $\Gamma^i_a$.

Finally, at the end of this procedure, we obtain the first class phase space
\begin{subequations}
\ba
G&\equiv&\partial_aE^a+A_a\time E^a\approx0,\\
C&\equiv&-\gamma^{-2}\frac{E^1\time E^2}{\det(\widetilde{E})}\cdot\Big(2\partial_1A_2+A_1\time A_2+(1-\gamma^2)(2\partial_1\Omega_2-\Omega_1\time\Omega_2)\Big)\approx0,\\
H_a&\equiv&-\gamma^{-1}E^a\cdot\big(\partial_1A_2-\partial_2A_1+A_1\time A_2\big)\approx0.
\ea
\end{subequations}
with the only non-vanishing Poisson bracket $\lb E^a_i,A^j_b\rb=-\gamma\delta^a_b\delta^j_i$, where
\be
A^i_a\equiv-\gamma\tilde{A}^i_a=(\Gamma^1_a-\gamma\omega^1_a,\Gamma^2_a-\gamma\omega^2_a,\Gamma^3_a-\gamma K^3_a)
\ee
is the three-dimensional Ashtekar-Barbero connection.

\section*{Conclusion}

\noindent In this note, we have tried to clarify at the classical level some issues related to the Barbero-Immirzi ambiguity. By performing a careful canonical analysis of the Holst action in the time gauge, we have re-derived the phase space structure of the Ashtekar-Barbero theory, and in particular identified at the Hamiltonian level the various components of the covariant torsion tensor. Since the phase space has an explicit dependency on $\gamma$, and this parameter disappears from the Lagrangian theory when the torsion-free equations of motion are resolved, it is natural to expect that in the Hamiltonian theory $\gamma$ will disappear when all the components \eqref{torsion components} of the torsion are vanishing. However, we have identified some subtleties in this reasoning. The first one is that when we work in the time gauge, some of the equations that correspond to the dynamics of $\chi$ just become fixation of Lagrange multipliers (i.e. \eqref{torsion00a no TG} becomes \eqref{torsion00a}). Then if \eqref{torsion00a} is not imposed in the quantum theory, it is to be expected that the dependency on $\gamma$ will survive, even at the dynamical level. Likewise, apart from $T^i_{ab}$ which vanishes classically because of the resolution of the second class constraints, the remaining torsion components involve the dynamics and the Gauss constraint, so even if they are imposed in the quantum theory they will necessarily vanish only on average, and the $\gamma$-dependency will not disappear. An even more problematic aspect is related to the transformation behavior of the Ashtekar-Barbero connection under time evolution. Indeed, since it does not transform as a spacetime connection, even if we focus only on the classical theory and impose strongly all the torsion-free equations, one can expect the theory to still be $\gamma$-dependent.

If one insists in performing the Hamiltonian quantization of a first order formulation of gravity, it is clear that the torsion-free condition will never be imposed strongly because some of its components involve dynamical equations. Now, as far as the Barbero-Immirzi ambiguity is concerned, this observation is not problematic if one quantizes a $\gamma$-independent symplectic structure such as the one obtained in the Lorentz-covariant formulation, and chooses to work with a true spacetime connection. However, if the parametrization of the phase space carries a $\gamma$-dependency, this latter will propagate in the quantum theory, and it seems not possible to expect the dynamics and the fixation of Lagrange multipliers to fully eliminate it.

This discussion should be put in the context of the following question: What is the meaning of the Barbero-Immirzi parameter? Is it simply a quantization ambiguity that results from an unfortunate (but nonetheless very useful!) parametrization of the phase space, or is it a fundamental dimensionless constant of quantum gravity? Recent developments \cite{usBH} are in fact pointing towards a picture in which $\gamma$ can be naturally removed by performing an analytic continuation to the self-dual value $\gamma=\pm i$, which was the original structure of the Ashtekar variables \cite{ashtekar}.

The three-dimensional model introduced in section \ref{sec:2} can certainly serve as a useful testbed to clarify all the subtleties related to the role of $\gamma$ in four-dimensional loop quantum gravity \cite{BGNY,BGNY2}. It would be interesting to study further the classical theory and compare it with the Chern-Simons formulation of three-dimensional gravity, which is known to be defined with a non-commutative connection. Another interesting aspect to investigate would be the description of the BTZ black hole in the presence of the three-dimensional Barbero-Immirzi parameter. Indeed, since there is now a loop quantum gravity description of the BTZ black hole entropy \cite{usBTZ}, it should in principle be possible to draw conclusions concerning the physical meaning of $\gamma$.

\section*{Acknowledgments}

\noindent We would like to thank Sergei Alexandrov for very useful comments and discussions. MG is supported by the NSF Grant PHY-1205388 and the Eberly research funds of The Pennsylvania State University.

\appendix

\section{Dirac bracket in the time gauge}
\label{appendix:Dirac}

\noindent An alternative to using the explicit resolution of second class constraints $S^i\approx0$ \eqref{secondclassS} and $\Psi^{ab}\approx0$ \eqref{Psiconstraints} that leads to the Levi-Civita connection (\ref{levi-civita solution}), is to work with the Dirac bracket. Its computation in the time gauge could be particularly interesting to establish a contact with the covariant formulation mainly developed by Alexandrov \cite{alexandrov1,alexandrov2,alexandrov3,alexandrov4}. Furthermore, by using the Dirac bracket instead of the explicit resolution $\omega^i_a=\Gamma^i_a(e)$, we can somehow keep at hand all the components of the connection, i.e. the rotational and the boost part, without singling out the rotational part by solving it in terms of the tetrad field.

To compute the bracket, it is convenient to use the freedom in choosing the parametrization of the second class constraints in order to replace the 3 constraints $S^i$ and the 6 constraints $\Psi^{ab}$ by the 9 equivalent constraints
\be
T^i_a\equiv\eps^{bc}_{~~a}(\partial_be^i_c-\eps^i_{~jk}\omega_b^je_c^k)\approx0.
\ee
The set of second class constraints is then given by $\pi^a_i\approx0$ and $T^i_a\approx0$, and we can compute the Dirac matrix along with its inverse:
\be
\Delta^{ij}_{ab}\equiv\lb\pi^a_i,T^j_b\rb=-\eps^{ij}_{~~k}\eps_{ab}^{~~c}e^k_c=\f{2}{\sqrt{\det(E)}}E^b_{[i}E^a_{j]},
\qquad\qquad
(\Delta^{-1})^{jk}_{bc}=\f{1}{2\det(e)}(2e^j_ce^k_b-e^k_ce^j_b).
\ee
Note that it is much simpler to compute the Dirac bracket with the constraints $T^i_a\approx0$ instead of using the original set.

The Dirac bracket between any two phase space functions $f$ and $g$ is defined as
\be
\lb f,g\rb_\text{D}=\lb f,g\rb-\lb f,\pi^a_i\rb(\Delta^{-1})^{ij}_{ab}\lb T^j_b,g\rb-\lb f,T^i_a\rb(\Delta^{-1})^{ij}_{ab}\lb\pi_j^b,g\rb.
\ee
Among all the canonical variables $E^a_i$, $A^i_a$, $\omega^i_a$, and $\pi^a_i$, the only non-trivial Dirac brackets are given by
\be
\lb E^a_i,A^j_b\rb_\text{D}=-\gamma\delta^a_b\delta^j_i,
\qquad\qquad
\lb\omega^i_a,A^j_b\rb_\text{D}=(\Delta^{-1})^{ik}_{ac}\lb T^k_c,A^j_b\rb.
\ee
A relevant question that we do not address here is wether this formalism allows to define new $\su(2)$ connections with interesting properties. This should be ultimately compared with the non-commutative shifted connection of the Lorentz-covariant formulation.

\section{Symmetry reduction of the four-dimensional Holst action}
\label{appendix symmetry red}

\noindent In this appendix, we show that the three-dimensional Holst action \eqref{3d action} can be obtained from a symmetry reduction of the four-dimensional Holst action. This result strongly supports the idea that our three-dimensional action is the lower-dimensional analogue of the Holst action.

The Holst action is given by \eqref{holst action}:
\be
S\equiv S_\text{HP}+\gamma^{-1}S_\text{H}=\frac{1}{4}\int_\mathcal{M}\de^4x\,\eps^{\mu\nu\rho\sigma}\left(\frac{1}{2}\eps_{IJKL}e_\mu^Ie_\nu^J 
F_{\rho\sigma}^{KL}+\gamma^{-1}\delta_{IJKL}e_\mu^Ie_\nu^JF_{\rho \sigma}^{IJ}\right).
\ee
In order not to reduce the internal gauge group $\SO(4)$, we only perform a space-time reduction. This can be done by assuming that the four-dimensional space-time has the topology $\mathcal{M}_4=\mathcal{M}_3\times\mathbb{I}$ where $\mathcal{M}_3$ is a three-dimensional space-time, and $\mathbb{I}$ is a spacelike segment with coordinates $x^3$. In this way, we single out the third spatial component $\mu=3$. Let us now impose the conditions
\be\label{symmetry conditions}
\partial_3=0,\qquad\qquad
\omega_3^{IJ}=0.
\ee
The first condition means that the fields do not depend on the third spatial direction $x^3$. The second one means that the parallel transport along the real line $\mathbb{I}$ is trivial. Therefore, the covariant derivative of the fields along the third direction vanishes.

Using the conditions \eqref{symmetry conditions}, a direct calculation shows that the four-dimensional Holst action reduces to
\be
S_\text{red}=-\int_\mathbb{I}\de x^3\int_{\mathcal{M}_3}\de^3x\,\eps^{\mu\nu\rho}\left(\f{1}{2}\eps_{IJKL}e_3^Ie^J_\mu F^{KL}_{\nu\rho}+\gamma^{-1}e_3^Ie^J_\mu F^{IJ}_{\nu\rho}\right),
\ee
where now $\mu,\nu,\rho$ are three-dimensional space-time indices, and $\de^3x=\de x^0\de x^1\de x^2$ is the volume form of $\mathcal{M}_3$. Apart from a global multiplicative factor that is not relevant at all, we recover the three-dimensional action \eqref{3d action} with the Barbero-Immirzi parameter, provided that we set $x^I\equiv e_3^I$.

\section{Details on the $\boldsymbol{\gamma}$-dependent three-dimensional gauge}
\label{appendix details 3d}

\noindent In this appendix, we give some intermediate expressions that were used to derive \eqref{3d gamma action}. It is possible to write \eqref{3d action 2} as the sum of a canonical term, a term involving $\omega_0$, a term involving $K_0$, and finally a term involving $e_0$.

\subsection{The canonical term}

\noindent It is easy to see that the canonical term is given by
\ba
&&-2\eps^{ab}\left[e_a\cdot\partial_0(\omega_b+\gamma^{-1}K_b)+(x\time e_a)\cdot\partial_0(K_b+\gamma^{-1}\omega_b)\right]\nonumber\\
&=&\widetilde{E}^a_i\partial_0(\omega^i_a+\gamma^{-1}K^i_a)+X^a_i\partial_0(K^i_a+\gamma^{-1}\omega^i_a)\nonumber\\
&=&E^a_i\partial_0\tilde{A}^i_a.
\ea

\subsection{The constraints $\boldsymbol{G_i}$ and $\boldsymbol{S_i}$}

\noindent The Gauss constraint $G_i$ and the second class constraint $S_i$ are given by the terms in \eqref{3d action 2} containing the variables $\omega_0$ and $K_0$. The term involving $\omega_0$ is
\ba
&&2\eps^{ab}\left[(x\time e_a+\gamma^{-1}e_a)\cdot(\omega_0\time K_b)+(e_a+\gamma^{-1}x\time e_a)\cdot(\partial_b\omega_0+\omega_0\time\omega_b)\right]\nonumber\\
&=&\omega_0\cdot\left[\partial_a(\widetilde{E}^a+\gamma^{-1}X^a)+(X^a+\gamma^{-1}\widetilde{E}^a)\time K_a+(\widetilde{E}^a+\gamma^{-1}X^a)\time\omega_a\right].
\ea
The term involving $K_0$ is
\ba
&&2\eps^{ab}\left[(x\time e_a+\gamma^{-1}e_a)\cdot(\partial_bK_0+K_0\time\omega_b)+(e_a+\gamma^{-1}x\time e_a)\cdot(K_0\time K_b)\right]\nonumber\\
&=&K_0\cdot\left[\partial_a(X^a+\gamma^{-1}\widetilde{E}^a)+(X^a+\gamma^{-1}\widetilde{E}^a)\time\omega_a+(\widetilde{E}^a+\gamma^{-1}X^a)\time K_a\right].
\ea
Introducing the index $\alpha=\{1,2\}$ to write $i=\{1,2,3\}=\{\alpha,3\}$, we can express the constraints imposed by the multipliers $\omega_0$ and $K_0$ as
\ba
&&\left\{
\begin{array}{l}\vspace{0.2cm}
\partial_a(\widetilde{E}^a+\gamma^{-1}X^a)+(X^a+\gamma^{-1}\widetilde{E}^a)\time K_a+(\widetilde{E}^a+\gamma^{-1}X^a)\time\omega_a\approx0\\\vspace{0.2cm}
\partial_a(X^a+\gamma^{-1}\widetilde{E}^a)+(X^a+\gamma^{-1}\widetilde{E}^a)\time\omega_a+(\widetilde{E}^a+\gamma^{-1}X^a)\time K_a\approx0
\end{array}\right.\\\vspace{0.2cm}
&\Leftrightarrow&\left\{
\begin{array}{l}\vspace{0.2cm}
\partial_a\widetilde{E}^a_\alpha+\gamma^{-1}\eps^{\alpha\beta}\widetilde{E}^a_\beta K^3_a-\eps^{\alpha\beta}X^a_3K^\beta_ a+\eps^{\alpha\beta}\widetilde{E}^a_\beta\omega^3_a-\gamma^{-1}\eps^{\alpha\beta}X^a_3\omega^\beta_a\approx0\\\vspace{0.2cm}
\gamma^{-1}\partial_aX^a_3+\gamma^{-1}\eps^{\alpha\beta}\widetilde{E}_\alpha K^\beta_a+\eps^{\alpha\beta}\widetilde{E}^a_\alpha\omega^\beta_a\approx0\\\vspace{0.2cm}
\gamma^{-1}\partial_a\widetilde{E}^a_\alpha+\gamma^{-1}\eps^{\alpha\beta}\widetilde{E}^a_\beta\omega^3_a-\eps^{\alpha\beta}X^a_3\omega^\beta_ a+\eps^{\alpha\beta}\widetilde{E}^a_\beta K^3_a-\gamma^{-1}\eps^{\alpha\beta}X^a_3K^\beta_a\approx0\\\vspace{0.2cm}
\partial_aX^a_3+\gamma^{-1}\eps^{\alpha\beta}\widetilde{E}_\alpha\omega^\beta_a+\eps^{\alpha\beta}\widetilde{E}^a_\alpha K^\beta_a\approx0
\end{array}\right.\\\vspace{0.2cm}
&\Leftrightarrow&\left\{
\begin{array}{l}\vspace{0.2cm}
\partial_a\widetilde{E}^a_\alpha+\eps^{\alpha\beta}(B^3_a\widetilde{E}^a_\beta-B^\beta_aX^a_3)\approx0\\\vspace{0.2cm}
\partial_aX^a_3+\gamma\eps^{\alpha\beta}\widetilde{E}^a_\alpha\tilde{A}^\beta_a\approx0\\\vspace{0.2cm}
\partial_a\widetilde{E}^a_\alpha+\gamma(\tilde{A}^3_a\widetilde{E}^a_\beta-\tilde{A}^\beta_aX^a_3)\approx0\\\vspace{0.2cm}
\partial_aX^a_3+\eps^{\alpha\beta}\widetilde{E}^a_\alpha B^\beta_a\approx0,
\end{array}\right.\label{system gauss}
\ea
where we have introduced the variable
\be
B^i_a=(B^\alpha_a,B^3_a)\equiv(K^\alpha_a+\gamma^{-1}\omega^\alpha_a,\omega^3_a+\gamma^{-1}K^3_a).
\ee
Now, one can see that the second and third equations of the system \eqref{system gauss} are equivalent to
\be\label{constraint intermed 1}
G\equiv\partial_aE^a-\gamma\tilde{A}_a\time E^a\approx0,
\ee
while the first and fourth equations are equivalent to
\be\label{constraint intermed 2}
\partial_aE^a-B_a\time E^a\approx0.
\ee
Finally, combining \eqref{constraint intermed 1} and \eqref{constraint intermed 2} and using the definition
\be\label{relation Omega B A}
\Omega^i_a\equiv\f{\gamma}{\gamma+1}(\tilde{A}^i_a+B^i_a),
\ee
we obtain the constraint $S\equiv\partial_aE^a-\Omega_a\time E^a\approx0$. Keeping track of the multipliers $\omega_0$ and $K_0$ during the above steps, it is easy to obtain the expressions \eqref{multipliers ab 3d}.

\subsection{The constraints $\widetilde{\boldsymbol{H}}^{\boldsymbol{a}}$ and $\boldsymbol{C}$}

\noindent The term involving $e_0$ is
\ba
&&\eps^{ab}e_0\cdot\Big\lbrace\partial_a(\omega_b+\gamma^{-1}K_b)-\partial_b(\omega_a+\gamma^{-1}K_a)+\omega_b\time\omega_a-K_a\time K_b-\gamma^{-1}K_a\time\omega_b+\gamma^{-1}K_b\time\omega_a\nonumber\\
&&+\Big(\partial_a(K_b+\gamma^{-1}\omega_b)-\partial_b(K_a+\gamma^{-1}\omega_a)-K_a\time\omega_b+K_b\time\omega_a+\gamma^{-1}\omega_b\time\omega_a-\gamma^{-1}K_a\time K_b\Big)\time x\Big\rbrace\nonumber\\
&\equiv&\eps^{ab}e_0\cdot H_{ab}.\label{3d hamiltonian gamma}
\ea
The components $i=\{\alpha,3\}$ of $H^i_{ab}$ are given by
\ba
H^\alpha_{ab}&=&\partial_a(\omega^\alpha_b+\gamma^{-1}K^\alpha_b)-\partial_b(\omega^\alpha_a+\gamma^{-1}K^\alpha_a)\nonumber\\
&&+\eps^{\alpha\beta}\Big(\omega^\beta_b\omega^3_a-\omega^3_b\omega^\beta_a-K^\beta_aK^3_b+K^3_aK^\beta_b
+\gamma^{-1}\left[K^3_a\omega^\beta_b-K^\beta_a\omega^3_b+K^\beta_b\omega^3_a-K^3_b\omega^\beta_a\right]\Big)\nonumber\\
&&-\eps^{\alpha\beta}x^\beta\Big\{\partial_a(K^3_b+\gamma^{-1}\omega^3_b)-\partial_b(K^3_a+\gamma^{-1}\omega^3_a)
+\eps_{\gamma\delta}\Big(K^\gamma_b\omega^\delta_a-K^\gamma_a\omega^\delta_b+\gamma^{-1}\omega^\gamma_b\omega^\delta_a-\gamma^{-1}K^\gamma_aK^\delta_b\Big)\Big\}\nonumber\\
&\equiv&M_{ab}^\alpha-\eps^{\alpha\beta}x^\beta M_{ab}^3,\label{way to Ha}
\ea
and
\ba
H^3_{ab}&=&\partial_a(\omega^3_b+\gamma^{-1}K^3_b)-\partial_b(\omega^3_a+\gamma^{-1}K^3_a)
+\eps_{\alpha\beta}\left(\omega^\alpha_b\omega^\beta_a-K^\alpha_aK^\beta_b-\gamma^{-1}K^\alpha_a\omega^\beta_b+\gamma^{-1}K^\alpha_b\omega^\beta_a\right)\nonumber\\
&&-\eps_{\alpha\beta}x^\alpha\Big\{\partial_a(K^\beta_b+\gamma^{-1}\omega^\beta_b)-\partial_b(K^\beta_a+\gamma^{-1}\omega^\beta_a)\nonumber\\
&&+\eps_{\beta\delta}\Big(-K^\delta_a\omega^3_b+K^3_a\omega^\delta_b+K^\delta_b\omega^3_a-K^3_b\omega^\delta_a
+\gamma^{-1}\left[\omega^\delta_b\omega^3_a-\omega^3_b\omega^\delta_a-K^\delta_aK^3_b+K^3_aK^\delta_b\right]\Big)\Big\}\nonumber\\
&\equiv&N_{ab}^3-\eps_{\alpha\beta}x^\alpha N_{ab}^\beta.\label{way to C}
\ea
Now, with \eqref{way to Ha} and \eqref{relation Omega B A}, one can rewrite the components $M^\alpha_{ab}$ and $M^3_{ab}$ in terms of $\tilde{A}^i_a$ and $B^i_a$ to see that we have
\ba
M^i_{ab}&=&\partial_a\tilde{A}^i_b-\partial_b\tilde{A}^i_a-\gamma\eps^i_{~jk}\tilde{A}^j_a\tilde{A}^k_b+\f{\gamma}{\gamma^2-1}\eps^i_{~jk}(\gamma\tilde{A}^j_a-B^j_a)(\gamma\tilde{A}^k_b-B^k_b)\nonumber\\
&=&\partial_a\tilde{A}^i_b-\partial_b\tilde{A}^i_a-\gamma\eps^i_{~jk}\tilde{A}^j_a\tilde{A}^k_b+(\gamma^{-1}-\gamma^{-3})\eps^i_{~jk}(\Omega^j_a-\gamma\tilde{A}^j_a)(\Omega^k_b-\gamma\tilde{A}^k_b).
\ea
Therefore, we can write that
\be\label{decomp H 1}
\widetilde{E}^a_\alpha H^\alpha_{12}=\widetilde{E}^a_\alpha M_{12}^\alpha+X^a_3M_{12}^3=E^a_iM_{12}^i\equiv\widetilde{H}^a.
\ee
Similarly, using \eqref{way to C} and \eqref{relation Omega B A}, one can see that
\ba
N^i_{ab}&=&\partial_aB^i_b-\partial_bB^i_a+\f{\gamma}{\gamma^2-1}\eps^i_{~jk}(B^j_a\tilde{A}^k_b+\tilde{A}^j_aB^k_b-\gamma\tilde{A}^j_a\tilde{A}^k_b-\gamma B^j_aB^k_b)\nonumber\\
&=&\gamma^{-1}\partial_a\tilde{A}^i_b-\gamma^{-1}\partial_b\tilde{A}^i_a+(1-\gamma^{-2})(\partial_a\Omega^i_b-\partial_b\Omega^i_a-\eps^i_{~jk}\Omega^j_a\Omega^k_b).
\ea
Using the expression
\ba
E^1\time E^2&=&(\widetilde{E}^1_1,\widetilde{E}^1_2,X^1_3)\time(\widetilde{E}^2_1,\widetilde{E}^2_2,X^2_3)\nonumber\\
&=&\big(\widetilde{E}^1_2X^2_3-\widetilde{E}^2_2X^1_3,\widetilde{E}^2_1X^1_3-\widetilde{E}^1_1X^2_3,\det(\widetilde{E})\big)\nonumber\\
&=&\det(\widetilde{E})(x_2,-x_1,1),
\ea
we can now write that
\be\label{decomp H 2}
H^3_{12}=N_{12}^3-\eps_{\alpha\beta}x^\alpha N_{12}^\beta=\f{(E^1\time E^2)^i}{\det(\widetilde{E})}N^i_{12}.
\ee
Finally, using \eqref{decomp H 1} and \eqref{decomp H 2} together with the definitions $e^\alpha_0\equiv N^a\widetilde{E}^a_\alpha/2$ and $e^3_0\equiv N/2$, we arrive at the decomposition
\be\label{decomposition H final}
2e^i_0H^i_{12}=2e^\alpha_0H^\alpha_{12}+2e^3_0H^3_{12}=N^aE^a_iM_{12}^i+N\f{(E^1\time E^2)^i}{\det(\widetilde{E})}N^i_{12}\equiv N^a\widetilde{H}_a+NC.
\ee

\end{document}